\documentclass[sigconf]{acmart}

\copyrightyear{2025}
\acmYear{2025}
\setcopyright{rightsretained}  
\acmConference[IUI '25]{30th International Conference on Intelligent User Interfaces}{March 24--27, 2025}{Cagliari, Italy}
\acmBooktitle{30th International Conference on Intelligent User Interfaces (IUI '25), March 24--27, 2025, Cagliari, Italy}
\acmDOI{10.1145/3708359.3712142}
\acmISBN{979-8-4007-1306-4/25/03}

\AtBeginDocument{%
  \providecommand\BibTeX{{%
    \normalfont B\kern-0.5em{\scshape i\kern-0.25em b}\kern-0.8em\TeX}}}

\usepackage{hyperref}
\usepackage{multirow}
\begin{document}

\title[Designing LLM-simulated Immersive Spaces to Enhance Autistic Children's Social Affordances Understanding]{Designing LLM-simulated Immersive Spaces for Autistic Children to Enhance Understanding of Social Affordances in Traffic Settings}


 \author{Yancheng Cao}
 \affiliation{%
   \institution{College of Design and Innovation, Tongji University \\ Institute for AI Industry Research, Tsinghua University}
   \city{Shanghai}
   \country{China}}
 \email{yanchengc7@outlook.com}

 \author{Yangyang HE}
 \affiliation{%
   \institution{Institute for AI Industry Research, Tsinghua University}
   \city{Beijing}
   \country{China}}
 \email{yangyang.he@gatech.edu}

  \author{Yonglin Chen}
 \affiliation{%
   \institution{Institute for AI Industry Research, Tsinghua University}
   \city{Beijing}
   \country{China}}
 \email{yonglin0711@gmail.com}

   \author{Menghan Chen}
 \affiliation{%
   \institution{Institute for AI Industry Research, Tsinghua University}
   \city{Beijing}
   \country{China}}
 \email{chenmenghan0815@163.com}

    \author{Shanhe You}
 \affiliation{%
   \institution{Institute for AI Industry Research, Tsinghua University}
   \city{Beijing}
   \country{China}}
 \email{moxdry@outlook.com}

    \author{Yulin Qiu}
 \affiliation{%
   \institution{Institute for AI Industry Research, Tsinghua University}
   \city{Beijing}
   \country{China}}
 \email{brantbaobao@outlook.com}

     \author{Min Liu}
 \affiliation{%
   \institution{Institute for AI Industry Research, Tsinghua University}
   \city{Beijing}
   \country{China}}
 \email{2660991881@qq.com}

     \author{Chuan Luo}
 \affiliation{%
   \institution{Tsinghua University}
   \city{Beijing}
   \country{China}}
 \email{chuanluobj@hotmail.com}

     \author{Chen Zheng}
 \affiliation{%
   \institution{Institute for AI Industry Research, Tsinghua University}
   \city{Beijing}
   \country{China}}
 \email{chen.zheng.psy@outlook.com}

      \author{Xin Tong}
 \affiliation{%
   \institution{Information Hub, HKUST(GZ)}
   \city{Guangzhou}
   \country{China}}
 \email{xint@hkust-gz.edu.cn}

\author{Jing Liang\textsuperscript{*}}
\affiliation{%
  \institution{College of Design and Innovation, Tongji University}
  \city{Shanghai} 
  \country{China}}
\email{12046@tongji.edu.cn}

\author{Jiangtao Gong\textsuperscript{*}}
\affiliation{%
  \institution{Tsinghua university}
  \city{Beijing}
  \country{China}}
\email{gongjiangtao@air.tsinghua.edu.cn}

\thanks{\textsuperscript{*}Mark corresponding authors.}

\renewcommand{\shortauthors}{Cao, et al.}

\begin{abstract}
 One of the key challenges faced by autistic children is understanding social affordances in complex environments, which further impacts their ability to respond appropriately to social signals. 
 In traffic scenarios, this impairment can even lead to safety concerns. In this paper, we introduce an LLM-simulated immersive projection environment designed to improve this ability in autistic children while ensuring their safety. We first propose 17 design considerations across four major categories, derived from a comprehensive review of previous research. Next, we developed a system called AIroad, which leverages LLMs to simulate drivers with varying social intents, expressed through explicit multimodal social signals. AIroad helps autistic children bridge the gap in recognizing the intentions behind behaviors and learning appropriate responses through various stimuli. A user study involving 14 participants demonstrated that this technology effectively engages autistic children and leads to significant improvements in their comprehension of social affordances in traffic scenarios. Additionally, parents reported high perceived usability of the system. These findings highlight the potential of combining LLM technology with immersive environments for the functional rehabilitation of autistic children in the future.

\end{abstract}







\maketitle

\section{Introduction}
Autism Spectrum Disorder (ASD) is one of the most complex and least understood conditions in the field of neurodiversity~\cite{muhle2004genetics}, and it is also among the most prevalent forms of neurodiversity~\cite{lorenz2017different}. Around one in every 36 children worldwide is diagnosed with ASD~\cite{maenner2023prevalence}. 
For these children, achieving social competence is challenging~\cite{hume2009increasing}.
Previous research has revealed that autistic children face difficulties in reading facial expressions~\cite{celani1999understanding, weeks1987salience}, maintaining eye contact~\cite{freeth2013affects,thorsson2024influence}, understanding gestures~\cite{attwood1988understanding, cairney2023interpretations}, and discerning vocal tones~\cite{brooks2013attention, schelinski2019relation}, among other areas. The social opportunities conveyed by these signals are referred to as \textit{social affordances}~\cite{loveland1991social}. 
The inappropriate understanding of \textit{social affordances} by autistic children further diminishes their opportunities for social integration~\cite{attwood2000strategies, chamberlain2007involvement} and consequently limit their chances to engage in and practice appropriate behaviors~\cite{lynch2009inclusive, smith2012developmental}, which are essential for functional rehabilitation~\cite{helt2008can, wood2019inclusive}.

Current social skills training programs for autistic children face several challenges. Firstly, these interventions often require the presence of other participants, which increases the complexity of the training process~\cite{bellini2007meta}. This is particularly problematic given that autistic children frequently struggle to form friendships in mainstream schools~\cite{rotheram2010social, carter2014promoting}, while the time allocated for group training sessions in special education programs is typically limited~\cite{kasari2011social}.
Moreover, existing programs often focus on specific skills without adequately explaining the underlying reasons for social cues, raising concerns about the generalizability of these interventions~\cite{porayska2012developing, grynszpan2014innovative}. Additionally, the backgrounds and narratives in training scenarios are predefined, which restricts their capacity to capture the complexity of real-life social interactions and to achieve a degree of sophistication for the agent~\cite{bernardini2014echoes}.
Compounding these issues is a notable lack of comprehensive synthesis regarding design considerations for interventions targeting autistic children~\cite{bozgeyikli2017survey}.

In the area of supporting and training social skills for autistic children, HCI community has made significant efforts, such as communication skill support~\cite{cha2021exploring, obiorah2021designing}, social trigger spaces~\cite{zanardi2022hoomie, wu2020squeeze}, emotion recognition and expression~\cite{tsai2021inclusion, lorenzo2016design, tang2024emoeden}, turn-taking and collaborative skills~\cite{bei2024starrescue}, among others~\cite{mosher2022immersive}.
In addition to support for these specific skills, previous research has also focused on particular contexts, such as the medical settings~\cite{guo2024dentar} and traffic scenarios~\cite{josman2008effectiveness}.
Education for autistic children in crossing scenarios primarily employs VR technology~\cite{saiano2015natural, liu2024virtual, goldsmith2008using, peng2019virtual, peng2019virtual}, along with touchable interfaces~\cite{singh2012gaming}. However, due to the complexity of social simulation, current immersive social affordance simulations for autistic children are still very limited, restricting the efficiency of training their social skills. Recently, the development of LLMs and their capabilities for social simulation present promising opportunities to address this issue \cite{paneru2024nexus, lyu2024designing, park2023generative, mishra2024towards}.

In this regard, this study aims to develop an LLM-simulated immersive space for autistic children that could facilitate social affordance understanding. 
In this paper, specifically, we hereby refer to \textbf{social affordance} as the range of \textbf{preferred behaviors opportunities as determined by the combination of explicit and implicit social signals~\cite{zheng2024putting, frith1994autism} and general social norms~\cite{ramstead2016cultural, tomasello2014natural}.}
In certain specific contexts, such as traffic scenarios, where pedestrians are required to understand social signals and make decisions~\cite{pele2017cultural}, this impairment can even pose risks to life and safety~\cite{curry2021comparison, falkmer2004transport}.
To ensure the system is acceptable and effective, we conducted a comprehensive literature review to synthesize design considerations for educational interventions targeting autistic children. 
Then, we use a street-crossing scenario as an example, as the ability to read social signals and understand social affordances becomes crucial for safety when traffic lights are absent~\cite{salamati2013event, vellenga2022driver}.
Towards this end, we leveraged LLMs to empower this space. Through the LLM's social simulation capabilities~\cite{park2023generative, pang2024self}, we generate a rich variety of social intentions based on different driving styles, each embodying distinct social affordances. These are then expressed through social signals in the immersive space using multimodal representations.
The system enables autistic children to interact safely with various car agents, allowing them to repeatedly practice, make mistakes, and receive corrective feedback.

Thus, the contributions of this paper are as follows: 
\begin{enumerate} 
\item Based on the comprehensive literature review, we identified four major categories of design considerations, encompassing a total of 17 of them regarding interactive educational technology for autistic children. 
\item We developed an immersive space with LLM simulation to facilitate autistic children's understanding of social affordances, allowing multimodal perception and output. 
\item The system can generate diverse driving intentions based on four different driving styles, with the underlying social affordances expressed through visual and auditory social signals.
\item A user experiment involving 14 autistic children was conducted to validate the system, demonstrating its usability and appeal to children, and confirming that it effectively helps them better choose the timing for crossing the street, which suggests an improved understanding of social affordances in traffic settings.
\end{enumerate}





\section{Related Work}


\subsection{Affordance and autistic children}
Affordances refer to the opportunities for perception and action that the environment offers to animals~\cite{gibson1977theory}. The term, coined by Gibson, evolved when Donald Norman introduced it to design, distinguishing between perceived and actual affordances~\cite{norman2013design, norman1988psychology} and highlighting potential everyday inconsistencies between them.
Loveland et al. categorize environmental affordances into three levels: physical transactions, cultural preferences, and social signals reflecting others' meanings. Autistic children often struggle with the latter two types~\cite{loveland1991social}. Difficulties with cultural affordances affect tool usage and may link to neurological issues~\cite{osiurak2017affordance}. Studies have explored interventions like electrical stimulation to enhance this understanding~\cite{lopes2015affordance++}. Understanding social signals involves grasping others' expectations, and research on helping autistic children in this area remains limited~\cite{ramstead2016cultural}.
Perceptual issues with affordances negatively impact social participation of autistic children and contribute to adverse behaviors, reducing learning opportunities~\cite{hellendoorn2014understanding}. While emotional signal transmission is crucial for interpersonal coordination~\cite{schmidt2008dynamics, hale2020you, hobson1991against}, autistic children's difficulties with social affordances~\cite{loveland1991social, hobson1993emotional, hobson1989sharing, hobson1990acquiring} lead to challenges in social adaptation and independent living~\cite{frith1994autism, schreibman1973overselective, kanner1972far, carter2005social}.

\subsection{Immersive environment for education}
Autistic children often experience significant deficits in social understanding and skills. Immersive systems provide effective learning environments and support mechanisms to help address these challenges ~\cite{cheng2015using}. Studies have shown these technologies successfully improve social behaviors and enhance communication and emotional skills ~\cite{halabi2017immersive,bekele2016multimodal,lorenzo2016design}. Through embodied interactions in virtual environments, children can safely explore new behavioral opportunities independently ~\cite{halabi2017immersive,matsentidou2014immersive}. These immersive settings provide a safe and inclusive space that reduces the hazards and unpredictability of real-life situations.
Immersive environments are typically delivered through VR systems with Head Mounted Displays (HMDs) or projection-based systems like Cave Automatic Virtual Environments (CAVE) ~\cite{bozgeyikli2017survey,burdea2003virtual}. While VR systems face limitations with autistic children often resisting headsets ~\cite{liu2017technology,soltiyeva2023my}, CAVE environments have proven effective for teaching safety skills, such as crossing streets or avoiding vehicles ~\cite{tzanavari2015effectiveness}. The current landscape of immersive environment systems presents significant research opportunities, particularly in integrating LLMs to enhance them with intelligent capabilities for memory, planning, and execution, though research in this area remains limited.

\subsection{LLM-simulated social interaction}
Large Language Models like ChatGPT has demonstrated remarkable ability in generating human-like responses \cite{brown2020language}. 
LLMs excel in fundamental tasks like translation \cite{susnjak2022chatgpt}, conversation generation, and code writing, and have made significant advances in more complex domains such as autonomous decision-making and role-playing, as demonstrated by applications such as AutoGPT \cite{yang2023autogpt} and HuggingGPT \cite{shen2023hugginggpt} in task planning and execution.
These advances have enabled practical applications across various domains - from creating character-aligned dialogues and simulating human behaviors in role-playing scenarios \cite{Shanahan2023, 9980408, park2023generative}, to serving specialized functions like educational teaching assistance \cite{Celik2022} and psychological counseling for individuals with high-functioning autism \cite{cho2023evaluating}.
Recent research has developed innovative LLM applications. A project created a simulated job fair environment for training generative agents with enhanced communication capabilities \cite{li2023metaagents}, while another advanced social network simulation by modeling agents with emotional and interactive capabilities \cite{gao2023s3}. Previous research introduced an alignment learning approach that leverages simulated society interactions, providing collective ratings and iterative feedback \cite{liu2023training}. These developments showcase LLMs' potential in mimicking human social interactions, suggesting a future where AI agents can participate in sophisticated social behaviors.

\begin{figure}[http]
  \includegraphics[width=0.4\textwidth]{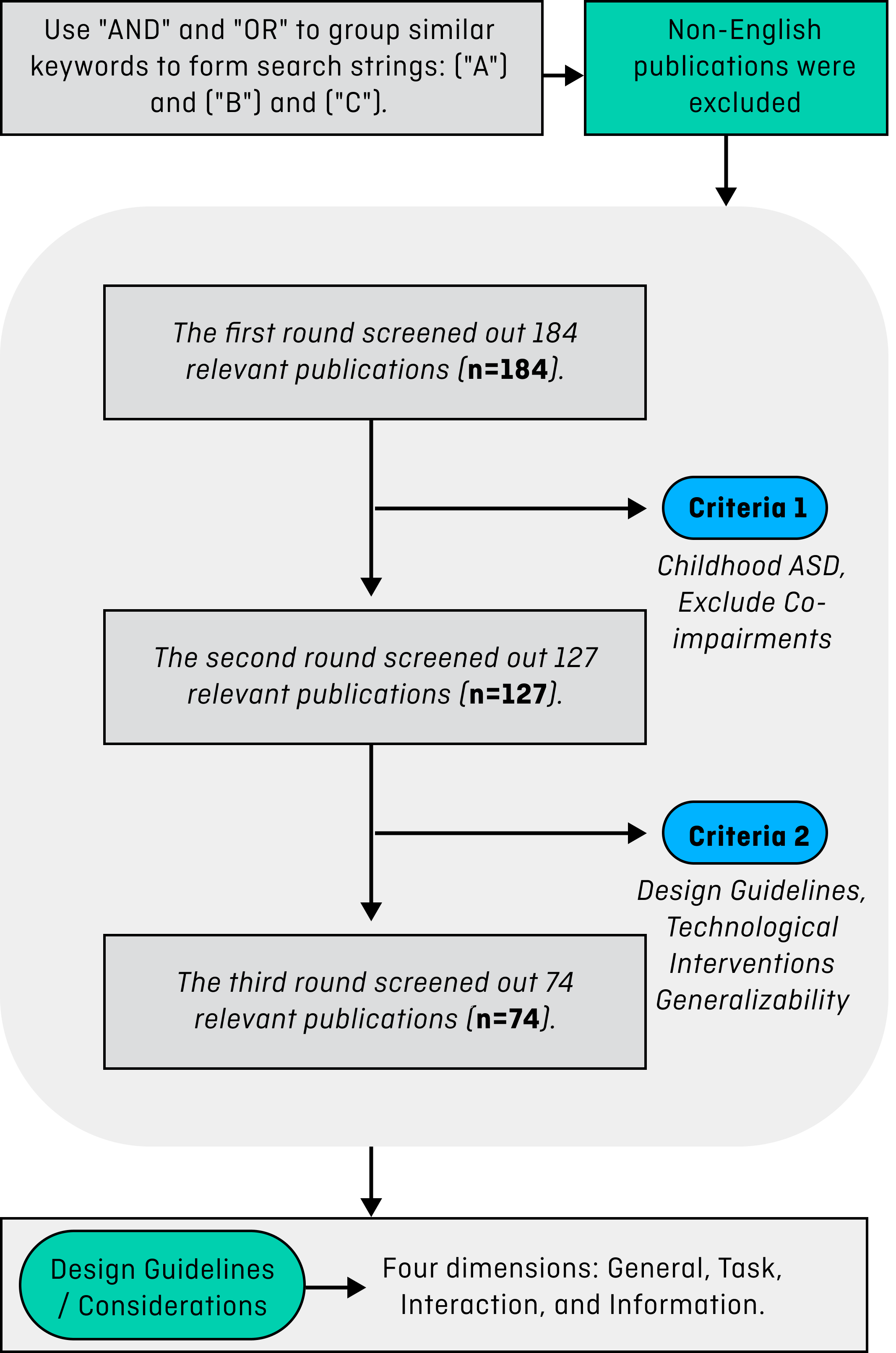}
  \caption{The literature screening procedure in this study involved two rounds of evaluation. This process resulted in the retention of 74 articles as the basis for design considerations.}
  \Description{This figure illustrates the literature screening process for the study. The literature screening procedure in this study involved two rounds of evaluation. This process resulted in the retention of 74 articles as the basis for design considerations.}
  \label{fig:Fig.1 The publication review procedure}
\end{figure}

\section{Design Consideration}

Design considerations for autism-focused systems remain limited in current literature, particularly regarding interaction-focused system design and LLM-enabled immersive spaces. To establish comprehensive guidelines for LLM-enabled immersive systems for autistic children, we conducted a systematic literature review (Fig. \ref{fig:Fig.1 The publication review procedure}).
Given LLM's nascent status and limited literature on its application for autistic children, our review includes studies involving other emergent technologies for this demographic, particularly focusing on immersive technologies. Our keywords cover three dimensions: (1) target audience - autistic children; (2) application domains - education, intervention, and rehabilitation; (3) associated technologies - virtual reality and similar systems. We used Boolean operators to create search strings like: ("autistic children" OR "Autistic children" OR "Autistic Spectrum Disorder") AND ("Education" OR "Intervention" OR "Rehabilitation") AND ("Artificial Intelligence" OR "LLM" OR "Immersive environment" OR "Virtual reality"). We searched six databases: IEEE Xplore, ACM Digital Library, SpringerLink, Elsevier, ScienceDirect, and Google Scholar.
Our team of five experts in engineering, psychology, and design conducted three screening rounds, reviewing titles, abstracts, and full texts. From 184 initial English articles, we applied two criteria: \textit{Criterion 1} excluded studies of non-childhood ASD cases and those with co-occurring cognitive impairments, while \textit{Criterion 2} identified articles addressing design considerations for technological interventions. 
The final 74 papers were analyzed to synthesize design guidelines across four dimensions: General, Task, Interaction, and Information.

\subsection{General Guidelines}

\subsubsection{Safe environment}
Autistic children often struggle to understand their surroundings, especially in terms of processing information, leading them to prefer environments that enhance their sense of control over interactions ~\cite{ke2016virtual,strickland2007evolution}.
Moreover, in the context of immersive environments, virtual reality devices such as head-mounted displays may induce negative experiences (e.g., dizziness or fatigue) for some autistic children ~\cite{liu2017technology,soltiyeva2023my}.

\textbullet \textit{General guideline 1: Provide an environment that makes children feel safe and comfortable.}

\subsubsection{Guidance and training}
It is essential to avoid lengthy and complex training sessions as autistic children often encounter difficulties in remembering and processing sequences ~\cite{grandin200228,davis2010guidelines}.
Given the attention comfort zone of autistic children, it is important to appropriately adjust the tone and pace of guidance appropriately and to be prepared to terminate training promptly if any discomfort arises ~\cite{van2008puzzling}.
Furthermore, studies on design considerations for autitic children emphasize the importance of providing instructions and the need for customized content adjustments tailored to the unique characteristics of this population ~\cite{dautenhahn2000design,millen2010development,barry2006interaction}.

\textbullet \textit{General guideline 2: Provide customized guidance and training in advance.}

\subsection{Task Guidelines}
\subsubsection{Objectives}
To facilitate participation for autistic children in current systems, it is crucial to establish a clear and explicit goal during task execution.~\cite{liu2017technology,whyte2015designing,bartoli2014designing}.
Bartoli et al. mentioned that the task objectives within the system should focus on a single task and a series of clear ,manageable actions (e.g., "swinging arms to hit an object"), to aid cognitive processes related to organizing actions, thereby achieving the set goals~\cite{bartoli2014designing}.

\textbullet \textit{Task guideline 1: Provide clear and understandable task objectives.}

\subsubsection{Repeatability and predictability}
Unpredictability may elicit a host of adverse reactions, such as anxiety, in autistic children, necessitating the use of repeatable tasks to facilitate learning and maintain stability ~\cite{liu2017technology,bartoli2014designing}.
System design should ensure that game tasks for autistic children are both repeatable and easily transitioned to higher levels to prevent discouragement. 
Whyte et al. emphasized the significance of repeatability and predictability, suggesting that repeating the same task can not only enhance mastery in autistic children but also provides the predictability they need, along with the anticipate future behaviors ~\cite{whyte2015designing}.

\textbullet \textit{Task guideline 2: Employ repeatable and predictable tasks, allowing children to attempt and practice repeatedly.}

\subsubsection{Storytelling}
The narrative describes a technique used in the design of systems for autistic children, where some research has utilized this technique alongside multimedia to encourage autistic children to complete tasks and create their own narratives ~\cite{hourcade2012multitouch,chatzara2014digital,kurniawan2018development}.
Certain story scenarios and cues can help improve the communication skills and symbolic functions of autistic children, thereby fostering their autonomy ~\cite{zaffke2015icanlearn}.
Soltiyeva et al. underscored the significance of storylines, noting that if autistic children become frightened by their surroundings, it can disrupt subsequent activities ~\cite{soltiyeva2023my}.
Hence, in spatial experiences, autistic children need to be guided and encouraged by stories from the outset. Moreover, since autistic children often struggle to transition between activities, the introduction of stories can assist them in understanding this transition. 

\textbullet \textit{Task guideline 3: Incorporate storylines to contextualize learning and enhance children's motivation.}

\subsubsection{Adjustable difficulty levels}
Given that each child exhibits distinct strengths and skill deficits, system design prioritize addressing each child's individual abilities and needs. This advocating for personalized adjustments for different children, requiring system that support a high level of personalization~\cite{liu2017technology,cai2023starrypia}.
System design should dynamically adjust the difficulty levels in alignment with the user's progression, and the intricacy of game tasks should be progressively increased~\cite{camargo2019designing,liu2017technology}.
Tasks should allowing for the introduction of incrementally more challenging tasks as children with ASD assimilate certain rules within the system. This approach concurrently facilitates the enhancement of both motor and cognitive complexities ~\cite{whyte2015designing,chua2017ict,bartoli2014designing,lee2012effects}.
As autistic children acquire and reinforce appropriate skills, they typically manifest an evolving spectrum of needs, such as escalating requirements for motor, cognitive, and social capabilities ~\cite{bartoli2014designing,lee2012effects}.
Furthermore, research in virtual environments has demonstrated that a gradually increment in task difficulty can yield superior outcomes for autistic children~\cite{strickland2007evolution,neale2002exploring,kee2012universal}.
Pivotal Response Treatment (PRT) further suggests that the emphasis of each task should adapt to individual progress, accommodating more advanced objectives and needs ~\cite{koegel2003empirically,mohammadzaheri2014randomized}.

\textbullet \textit{Task guideline 4: Implement adaptable game difficulty levels, judiciously escalating complexity in correspondence with the personal capabilities of each individual child.}

\subsection{Interaction Guidelines}
\subsubsection{Multimodal input}
Multimodal systems that integrate interaction through voice, writing, touch, and other means offer distinct advantages over unimodal systems by leveraging redundancy and complementarity. This facilitates autistic children in conveying their interactive inputs to the system across multiple dimensions ~\cite{vieira2017tell}.
Previous research has also demonstrated the effectiveness of multimodal interventions in enhancing the learning and social communication skills of autistic children~\cite{brady2015investigating,beaumont2013multimodal}.

\textbullet \textit{Interaction Guideline 1: Utilize multimodal perception at the input stage to collect behavioral data from the children.}

\subsubsection{Real-time feedback}
In interfaces designed for autistic children, feedback and signals related to the structure of interactive elements serve a crucial role in reinforcing the expected tasks for these children~\cite{konstantinidis2009using,davis2010guidelines}.
Real-time feedback, in particular, aligns closely with their interaction expectations. Previous research has observed that positive real-time feedback can significantly motivate participants and influence their engagement in subsequent interactions~\cite{konstantinidis2009using,lanyi2004multimedia,parsons2013chooses}.

\textbullet \textit{Interaction Guideline 2: Provide real-time feedback for each behavior exhibited by the children.}

\subsubsection{Reward mechanisms}
In the domain of system design for autistic children, providing positive reinforcement along with rewards is a strategy known as reward-based intervention ~\cite{liu2017technology,whyte2015designing}.
Bartoli et al. noted that offering reward stimuli to autistic children upon the completing tasks that meet expectations can enhance motivation and engagement and may implicitly augment their skills \cite{bartoli2014designing}. 
Research indicates that some autistic children, particularly those with moderate to low functioning, may not value rewards in the form of quantitative performance outcomes (e.g., scores) to the same extent as they are intrinsically interesting, such as videos or audio effects (e.g., entertaining animations or joyful music) ~\cite{bartoli2014designing,uzuegbunam2017mebook}.
Incorporating reward reinforcement in serious games designed for autistic children is an effective strategy that can enhance a variety of behaviors and skills ~\cite{wong2015evidence,bartoli2014designing,de2014daybyday}.

\textbullet \textit{Interaction Guideline 3: Provide positive reinforcement integrated with reward mechanisms.}

\subsubsection{Ease of input}
In system designs tailored for autistic children, simplicity and intuitiveness in control mechanisms are considered as superior design principles ~\cite{bozgeyikli2017survey}.
Complex and cumbersome input methods have been identified as sources of frustration for these children~\cite{davis2010guidelines,strickland2007evolution,kee2012universal,lanyi2004multimedia}. 
Redundant controls can cause confusion among autistic children, who may be more prone to abandonment and disengagement than typical users ~\cite{davis2010guidelines,wickramasinghe2020trustworthy}.

\textbullet \textit{Interaction Guideline 4: Leverage simple and intuitive input methods to facilitate children's ease of use.}

\subsubsection{Devices}
For autistic individuals who may experience motor difficulties, it is advisable to employ a limited number of input devices ~\cite{davis2010guidelines,mei2014usability,bozgeyikli2017survey}.
Previous research has suggested that reducing the number of control buttons within input devices can benefit autistic children
~\cite{davis2010guidelines,strickland2007evolution,kee2012universal}.
Moreover, autistic children may experience panic and resistance when using VR devices, with reports indicating that autistic children with anxiety disorders often fear and refuse to wear VR headsets ~\cite{soltiyeva2023my}.
Additionally, unfriendly methods of device usage can create barriers that hinder the normal progression of research activities. 

\textbullet \textit{Interaction Guideline 5: Utilize fewer devices that are also child-friendly to facilitate user input.}

\begin{table*}[htbp]
  \caption{Design Guidelines for designing immersive interactive system for autistic children.}
  \label{tab:design considertion}
  \resizebox{\textwidth}{!}{
    \begin{tabular}{lll}
      \toprule
      \multicolumn{2}{l}{\textbf{General Guidelines}} \\
      \midrule
      Provide an environment that makes children feel safe and in control. & ~\cite{grandin200228,davis2010guidelines,ke2016virtual,strickland2007evolution,liu2017technology,soltiyeva2023my} \\
      Provide customized guidance and training. & ~\cite{grandin200228,davis2010guidelines,van2008puzzling,dautenhahn2000design,millen2010development,barry2006interaction} \\
      \textit{} &  \\ 
      \midrule
      
      \multicolumn{2}{l}{\textbf{Task Guidelines}} \\
      \midrule
      Provide clear and understandable task objectives. & ~\cite{liu2017technology,whyte2015designing,bartoli2014designing} \\
      Employ repeatable and predictable tasks, allowing children to attempt and practice repeatedly. & ~\cite{liu2017technology,bartoli2014designing,whyte2015designing} \\
      Incorporate storylines to contextualize learning and enhance children's motivation. & ~\cite{hourcade2012multitouch,chatzara2014digital,kurniawan2018development,zaffke2015icanlearn,soltiyeva2023my} \\
      Implement adaptable game difficulty levels, judiciously escalating complexity in correspondence with the personal capabilities of each individual child. & ~\cite{camargo2019designing,liu2017technology,chua2017ict,bartoli2014designing,lee2012effects,whyte2015designing,strickland2007evolution,neale2002exploring,kee2012universal,koegel2003empirically,mohammadzaheri2014randomized,liu2017technology,cai2023starrypia} \\
      \textit{} &  \\ 
      \midrule
      
      \multicolumn{2}{l}{\textbf{Interaction Guidelines}} \\
      \midrule
      Utilize multimodal perception at the input stage to collect behavioral data from the children. & ~\cite{vieira2017tell,brady2015investigating,beaumont2013multimodal} \\
      Provide real-time feedback for each behavior exhibited by the children. & ~\cite{konstantinidis2009using,davis2010guidelines,lanyi2004multimedia,parsons2013chooses} \\
      Provide positive reinforcement integrated with reward mechanisms. & ~\cite{liu2017technology,whyte2015designing,bartoli2014designing,uzuegbunam2017mebook,uzuegbunam2017mebook} \\
      Leverage simple and intuitive input methods to facilitate children's ease of use. & ~\cite{bozgeyikli2017survey,davis2010guidelines,strickland2007evolution,kee2012universal,lanyi2004multimedia,wickramasinghe2020trustworthy} \\
      Utilize fewer devices that are also child-friendly to facilitate user input. & ~\cite{davis2010guidelines,mei2014usability,bozgeyikli2017survey,strickland2007evolution,kee2012universal} \\
      \textit{} &  \\ 
      \midrule
      
      \multicolumn{2}{l}{\textbf{Information Guidelines}} \\
      \midrule
      Employ vibrant colors and a consistent style. & ~\cite{grandgeorge2016atypical,adjorlu2020co,quill1997instructional,chua2017ict,grandgeorge2016atypical,quill1997instructional,chua2017ict} \\
      Implement simplified graphics, sound, and text. & ~\cite{chua2017ict,bozgeyikli2017survey,parsons2005adolescents,parsons2006virtual,fabri2005use,bartoli2014designing} \\
      Utilize cartoon characters as virtual avatars instead of realistic representations. & ~\cite{cai2023starrypia,rosset2008typical,robain2022measuring,zheng2017toon,liu2017technology,hsu2017investigating,zhao2020virtual} \\
      Provide appropriate dynamic stimuli in the scenes to capture children's attention. & ~\cite{liu2017technology,whyte2015designing,bartoli2014designing} \\
      Provide prompts for children using various forms of information presentation, including text, sound, and animation. & ~\cite{liu2017technology,chen2022excessive,hourcade2012multitouch,chatzara2014digital,kurniawan2018development,whyte2015designing,adjorlu2020co} \\
      Avoid the occurrence of overly unexpected stimuli in the scenes. & ~\cite{soltiyeva2023my,hourcade2012multitouch,bartoli2014designing,grandin200228,davis2010guidelines,van2008puzzling} \\
      \bottomrule
    \end{tabular}
  }
\end{table*}

\subsection{Information Guidelines}
\subsubsection{Vibrant colors and consistent style}
Autitstic Children often process and remember visual information better than verbal information and are more sensitive to visual sensory stimuli than typically developing children ~\cite{grandgeorge2016atypical,adjorlu2020co,quill1997instructional,chua2017ict}.
Research has suggested that systems designed for autistic children should incorporate more vibrant colors in the visual design to cater to this strength ~\cite{camargo2019designing,zhang2018design,alvarado2017valpodijo,yan2011sunny,siti2011edutism}.
Moreover, a consistent color style can provide comfort and help prevent overstimulation. 

 \textbullet \textit{Information Guideline 1: Employ vibrant colors and a consistent style.}

\subsubsection{Simplified information}
When designing graphics, sound, and text for systems aimed at autistic children, it is advised to avoid complex elements and lengthy compositions, as these may lead to distractions and, in severe cases, sensory overload ~\cite{chua2017ict,bozgeyikli2017survey,parsons2005adolescents,parsons2006virtual,fabri2005use}.
Bartoli et al. noted that excessive visual stimuli can cause anxiety in autisitic children because they may struggle to differentiate and interpret individual elements within a group, while too many auditory stimuli can create additional stress ~\cite{bartoli2014designing}.
Moreover, research has highlighted that simplified graphics can enhance information processing for autistic children~\cite{parsons2005adolescents,parsons2006virtual,fabri2005use}.


\textbullet \textit{Information Guideline 2: Implement simplified graphics, sound, and text.}

\subsubsection{Cartoon characters}
Some studies have indicated that autistic children are more visually attentive to and engage in eye contact with cute and stylized cartoon characters than with images of real people ~\cite{cai2023starrypia,rosset2008typical,robain2022measuring,zheng2017toon}.
Furthermore, the use of simple 2D cartoon-like avatars in system has been suggested as a more widely acceptable approach by autistic children compared to realistic human representations or complex characters ~\cite{liu2017technology,hsu2017investigating,zhao2020virtual}.

\textbullet \textit{Information Guideline 3: Utilize cartoon characters as virtual avatars instead of realistic representations.}

\subsubsection{Dynamic stimuli}
Prolonged static scenes in a system can result in a loss of attention in children, and in the case of autistic children, such static scenarios can even trigger motor rigidity ~\cite{liu2017technology,whyte2015designing}.
To counteract this, it is important to provide appropriate dynamic stimulation throughout the interaction to prevent the emergence of repetitive behaviors or motor rigidity in autistic children ~\cite{bartoli2014designing}.

\textbullet \textit{Information Guideline 4: Provide appropriate dynamic stimuli in the scenes to capture children's attention.}

\subsubsection{Multimodal Prompts}
Autistic children often have difficulty processing external stimuli, which can affect their awareness of how their actions impact others and the environment. To address this, they may require multiple forms of prompts to guide subsequent actions ~\cite{liu2017technology,chen2022excessive}.
Previous studies have utilized a combination of media in the design of systems for autistic children to encourage interaction and promote autonomy ~\cite{hourcade2012multitouch,chatzara2014digital,kurniawan2018development}.
Additionally, some research has integrated text, auditory, and visual cues to convey game instructions,  with visual aids designed based on expert advice to assist autistic children understand verbal commands ~\cite{liu2017technology,whyte2015designing,adjorlu2020co}.

\textbullet \textit{Information Guideline 5: Provide prompts for children using various forms of information presentation, including text, sound, and visual cues.}

\subsubsection{Avoid overly sudden stimuli}
Some autistic children exhibit atypical fears of loud noises and fast-moving objects, leading to recommendations for excluding such elements from system designs intended for them~\cite{soltiyeva2023my,hourcade2012multitouch,bartoli2014designing}.
Moreover, several studies have advised against incorporating sudden noises and unexpected, abrupt visual changes in the systems or educational content used by autistic children~\cite{grandin200228,davis2010guidelines,van2008puzzling}.

 \textbullet \textit{Information Guideline 6: Avoid the occurrence of overly unexpected stimuli in the scenes.}

\section{System Design of AIRoad}
In traffic scenarios without explicit traffic lights, understanding social affordances becomes crucial \cite{ramstead2016cultural, loveland1991social}. To address this need, we developed AIRoad, a system specifically designed for street-crossing situations. This section first presents AIRoad's conceptual design and demonstrates its application through concrete examples. We then show how these implementations align with our previously established design guidelines, followed by a detailed discussion of the system's software and hardware components.


\subsection{Concept Design of AIRoad: Bridging Behavior and Intentions to Enhance Understanding of Social Affordance}
\begin{figure*}[htbp]
    \centering
    \includegraphics[width=1.0\linewidth]{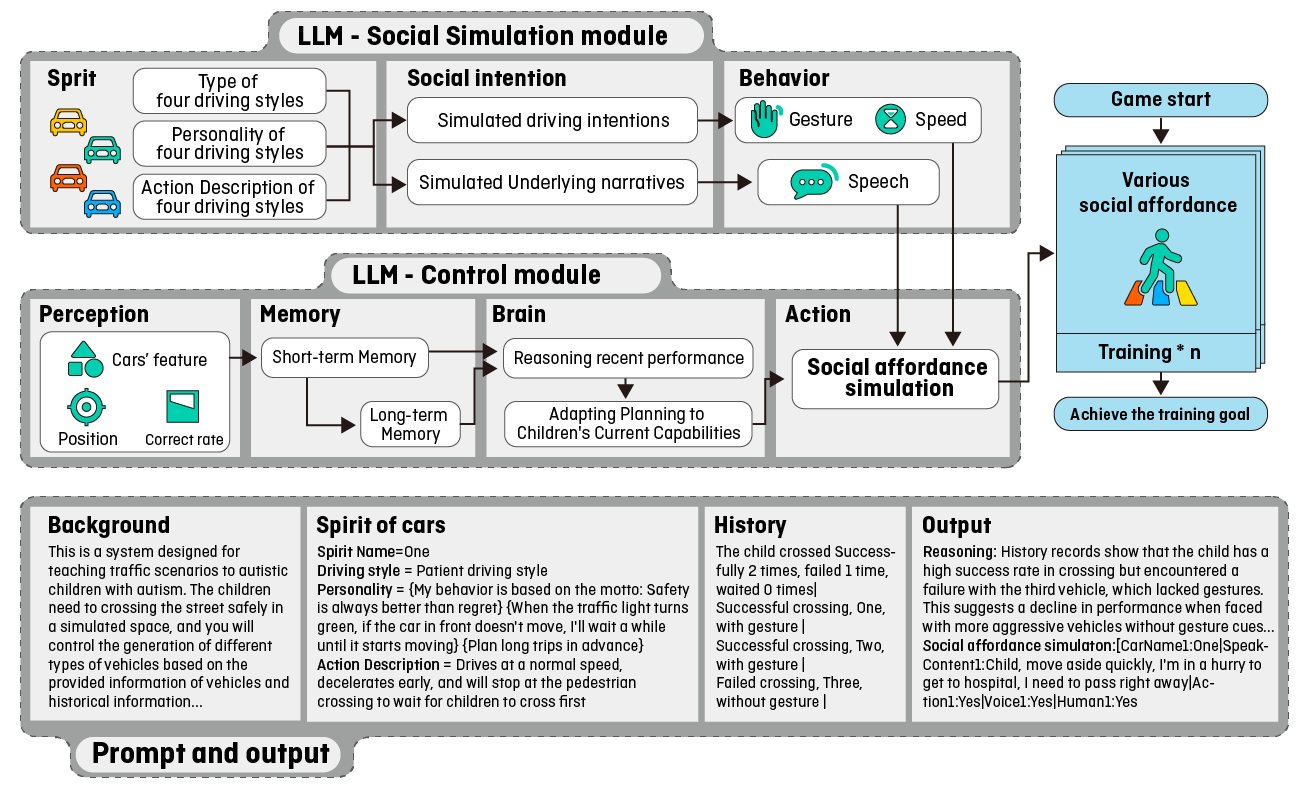}
    \caption{System framework of AIRoad. The social affordance simulation conducted by the LLM is detailed in the social simulation module. The control of game rules by the LLM is outlined in the control module. Key prompts and outputs related to the LLM are also displayed below.}
    \Description{This figure displays the system framework of AIRoad and the entire training process, including the adaptive adjustment and content generation by the LLM. In our work, we have developed an AI agent system based on LLM, which comprises an social simulation module and a control module.}
    \label{fig:System framework}
\end{figure*}

While neurotypical children naturally learn social signals through daily exposure~\cite{aboud2003developmental}, autistic children often require structured practice~\cite{helt2008can}. We developed a virtual street-crossing scenario enabling safe, controlled practice~\cite{shahmoradi2022cognitive}, as illustrated in Fig.~\ref{fig:System framework}.
AIRoad presents vehicle behaviors and their intentions through multiple modalities, helping autistic children bridge behavior-intention connections in assessing driver yielding behavior. The system creates an LLM-simulated environment featuring street-crossing scenarios without traffic lights, displaying vehicles with various driving intentions through multi-faceted projections.
Our driving intention generation is based on four styles from O. Taubman-Ben-Ari et al.~\cite{taubman2004multidimensional}: dissociative, anxious, risky, and patient driving. Vehicle intentions are expressed through behavioral cues (speed and gestures) indicating yielding decisions, with LLM-generated narratives conveyed through speech synthesis.
The LLM selects vehicle animations and generates corresponding intentions using literature-based prompts. Participants observe visual cues while hearing drivers' intentions through speech synthesis. During training, participants collect stars across road locations, requiring continuous interpretation of social affordances through multimodal information.
This immersive space provides autistic children with a controlled learning environment~\cite{shahmoradi2022cognitive}. Learning begins with enhanced scaffolding, while the system monitors performance to adjust difficulty and support levels. The LLM-powered adaptive space optimizes difficulty levels and provides multimodal scaffolding to facilitate skill development, ultimately enabling autistic children to comprehend social affordances effectively.

\subsection{Example Scenarios to Illustrate Specific Functions}
The system's core functions are illustrated through two contrasting examples. In the patient driving style, characterized by "Better safe than sorry," the vehicle slows down well in advance, and when participants show high error rates, the system displays a gesture encouraging pedestrians to cross. The LLM generates a matching narrative: "I'm heading to the supermarket to pick up a few things; there's no rush," which is conveyed through audio. Conversely, in the dissociative driving style, where the driver is "lost in thoughts or distracted," the vehicle maintains speed through the crossing without yielding. For participants with low error rates, no additional gesture cues are provided, while the LLM narrates: "I was on the phone and didn't see anyone on the road." These examples demonstrate how the system combines visual behaviors, adaptive gesture cues, and verbal intentions to help participants recognize different social affordances in crossing scenarios.

\subsection{System Development Under Design Guidelines}
The system development followed the aforementioned design considerations, denoted as DG1-DG17 based on their order in the table.

\subsubsection{\textbf{System Development Under General Guidelines}} 
To meet \textbf{DG1}, the system enables users to input a nickname, facilitating personalized addressing in the generated voice prompts. We selected a simple intersection common in everyday environments as the context for our study. To address \textbf{DG2}, an introductory page was developed to facilitate interactions, allowing children to engage in dialogue with various vehicles and trees, thereby familiarizing themselves with the spatial layout before their experience with AIRoad. During the experimental phase, the facilitator provided a comprehensive overview of the game tasks, which is elaborated upon in the subsequent section.

\subsubsection{\textbf{System Development Under Task Guidelines}} 
To align with \textbf{DG3}, we designed a clear and engaging objective: participants must collect stars while safely crossing the road, with emphasis on both speed and safety. In addressing \textbf{DG5}, we leveraged LLM's social simulation capabilities to generate diverse driving intention narratives, presenting them as engaging stories that enhance autistic children's engagement and understanding.
Following \textbf{DG4}, the system implements repeated road-crossing scenarios, allowing participants to progressively improve their ability to interpret vehicle driving intentions while learning from their mistakes. To satisfy For \textbf{DG6}, the system tracks participant errors through short-term and long-term memory, allowing the LLM to adjust difficulty levels dynamically. These adjustments include varying levels of cognitive support (such as gesture cues and voice-over narratives) and environmental challenges (like vehicle speeds and pedestrian interactions).

\subsubsection{\textbf{System Development Under Interaction Guidelines}} 

To meet \textbf{DG7} and \textbf{DG11}, AIRoad utilizes a portable Vive tracker and a lightweight microphone as input devices, facilitating multimodal input while ensuring ease of transport. To address \textbf{DG10}, we minimized the operational tasks required of participants; the tracker functions without intervention, and the microphone can be activated with a single button press. To fulfill \textbf{DG8}, all feedback animations are based on real-time position recognition. In accordance with \textbf{DG9}, the screen provides live updates on the number of stars collected and those yet to be obtained. Additionally, prior to the start of the experiment, participants were informed about the rewards system, which included sticker rewards distributed throughout the process.

\subsubsection{\textbf{System Development Under Information Guidelines}} 
To meet \textbf{DG12} and \textbf{DG14}, the system's visual design incorporates a cartoonish style for vehicles and environments, utilizing vibrant colors that are appealing (as illustrated in Fig. \ref{fig:cave space}). To address \textbf{DG13}, all graphics are designed to avoid excessive detail, and we have specified in the LLM prompts that generated speech should not exceed 25 words. To fulfill \textbf{DG15}, we created dynamic animations for moving vehicles, talking trees, collectible stars, and interfering pedestrians to stimulate engagement among autistic children. To meet \textbf{DG16}, the immersive space conveys prompts through multimodal elements such as vehicle animations, collision sounds, and narrative dialogues. Lastly, to \textbf{address DG17}, the vehicles encountered during training are positioned in fixed locations, and the voice prompts are generated using gentle tones to minimize sudden stimuli for autistic children.

\begin{figure*}[htbp]
\centering
   \includegraphics[width=0.9\textwidth]{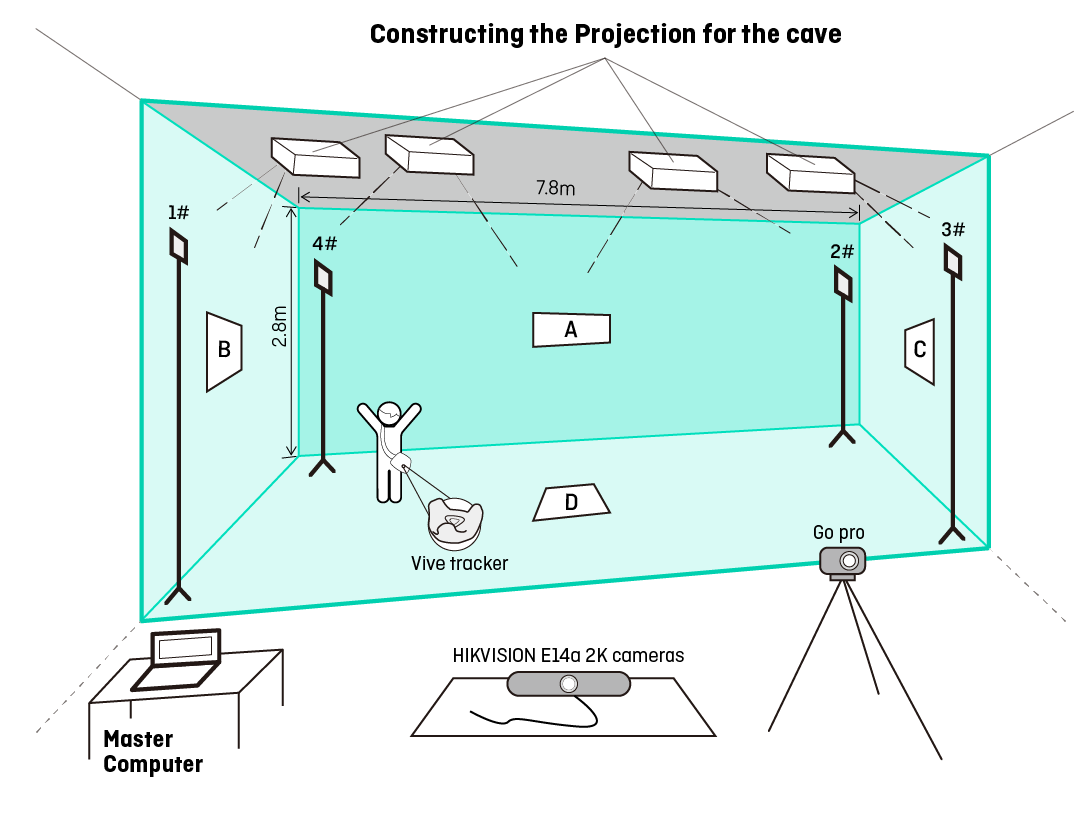}
   \caption{AIRoad is an AI-enabled immersive educational space tailored for autistic children. It facilitates autistic children's learning about social affordances in complex traffic scenarios through the construction of virtual transportation settings.}
   \Description{This figure shows the physical setup of AIRoad, which is located in a CAVE space consisting of four screens labeled A/B/C/D, with four projectors above. One HTC Vive base station is placed in each corner of the space. Users in the cave are required to wear a cross-body bag equipped with an HTC Vive tracker. Two cameras are set up in front of the CAVE space to record the user experience for analysis.}
   \label{fig:teaserCave}
\end{figure*}

\begin{figure*}[htbp]
  \includegraphics[width=1.0\textwidth]{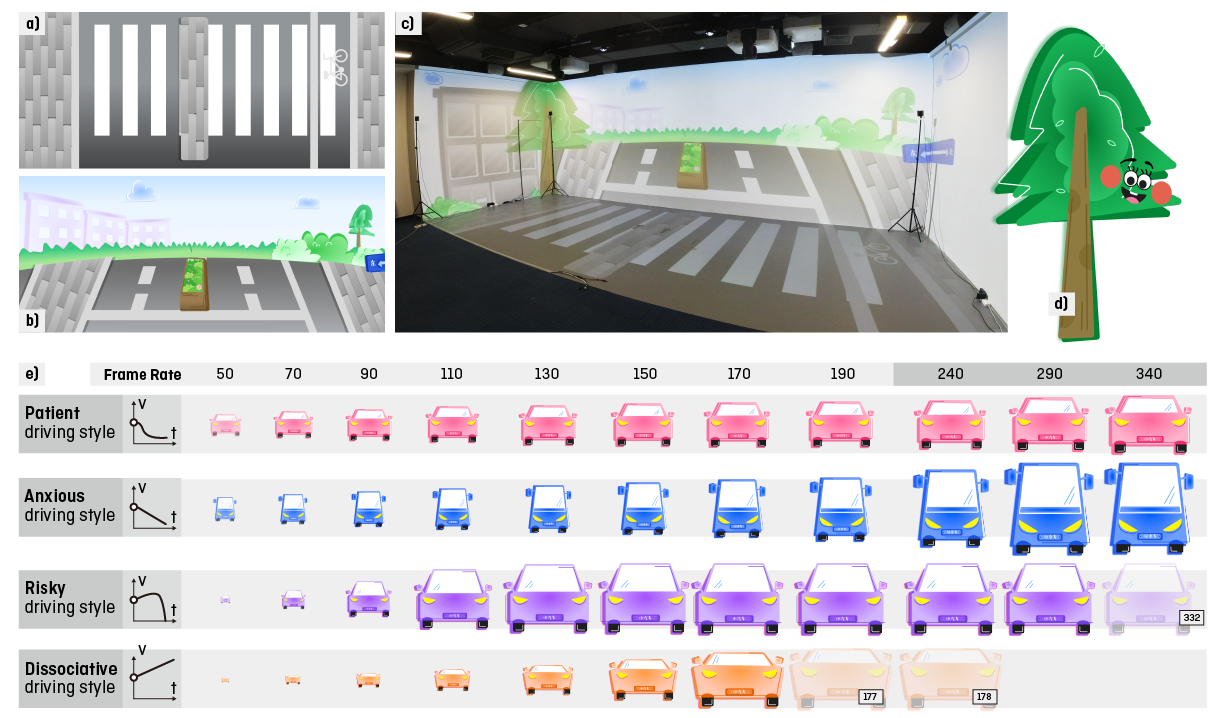}
  \caption{a) Projection on the ground; b) Projection on the wall; c) Overall real-life overview; d) Other cartoon elements; e) Creation of animation effects and vehicle driving style (30 frames per second).}
  \Description{This series of images is a visual presentation of AIRoad, featuring a) Projection on the ground; b) Projection on the wall; c) Overall real-life overview; d) Other cartoon elements; e) Creation of animation effects and vehicle driving style (30 frames per second).}
  \label{fig:cave space}
\end{figure*}

\subsection{System implementation}
We developed a Unity-based virtual environment with LLM-controlled assets, featuring safety zones and dual crosswalks. The space is equipped with four VR stations that track users' real-time positions to evaluate their behavior during simulated vehicle crossing scenarios. This immersive environment captures user behaviors and transmits logs to LLMs, which then generate standardized text formats to orchestrate traffic simulations using predefined Unity assets. To enhance acceptance among autistic children, we chose projection as the primary medium, with detailed rationale provided in Table.\ref{tab:design considertion}.

\subsubsection{Hardware and space setting}
As shown in Fig. \ref{fig:teaserCave}, AIRoad creates an immersive experience using a four-projection system with HTC Vive base stations mounted at the four upper corners, and the HTC Vive trackers are used to track the player's real-time position in virtual space. Notably, to avoid children participants feeling uncomfortable with wearing trackers, we strapped the tracker to a backpack and designed with a playful appearance, making them appealing and acceptable to most children so that they are willing to carry a backpack while playing.
Fig. \ref{fig:teaserCave} show that the projections displayed in faces \textit{A} and \textit{D} are of equal size (7.8 m in length and 2.8 m in width), while faces \textit{B} and \textit{C} are also of equal size (2.8 m in length and 2.8 m in width).
Additionally, the experimenter uses a master computer to remotely control the computer connected to the projection system and has set up a \textit{GoPro} and a \textit{HIKVISION E14a 2K} camera to record the experimental process.

\subsubsection{Technical framework}

AIRoad utilizes a prompt-driven decision-making approach for LLMs, leveraging prompts to describe scenarios and requirements, thereby facilitating the decision-making and thought processes of the LLMs. The related source code and associated prompts are available at \url{https://github.com/Minadocyc/AIRoad.git}

\textbf{LLM Model Selection.} GPT-4 Turbo was selected as the underlying model after comparative testing with other models such as GPT-3.5 and ERNIE. At the time of development, it was OpenAI's most capable model. It demonstrated superior performance in instruction compliance, social intention generation, and response time - particularly important given the lengthy prompts used.

\textbf{Prompt Development.} Our system interfaces with GPT-4 Turbo through OpenAI's API, utilizing multiple locally stored prompt files for system configuration. These files include Background (system context); Tool (available tools and output format standards); Social (detailing characteristics of four vehicle types); and History (performance history and recent logs). During operation, these files are consolidated into a comprehensive prompt, sent to the LLM for decision analysis, which then generates simulated traffic scenarios.

\textbf{Memory Module.} The memory module stores log information that encapsulates the entire scenario context, also known as the context of scene changes. Changes in player positions and scene content trigger updates that are logged into the Memory Module. This comprehensive context, provided by the Memory Module, ensures that the LLMs have all the necessary information at their disposal when making decisions. When LLMs are called upon to make decisions, a concatenation of all logs and prompts serves as the input for the LLMs.

\textbf{Spirit.} A spirit is an object that can execute commands, produce animation effects, and generate unique sounds. In our scenario, the setting of it through personified characteristics is defined by adjustable JSON prompts. Each spirit has customizable attributes such as type, personality, position, responsibilities, actions, and voice settings. For example, a pink toy car is characterized as cute and lively, assisting children in crossing the street safely. Its voice is generated using a trained VITS model, employing a gentle female voice to convert text to speech with a distinct tonal quality.


\textbf{Real-time feedback.} Scene dynamics include participant positions, traffic light states, and spirit behavior. Due to LLMs' limitations with token, the scene is divided into areas, and each area will have a specific text description for functions. with player movements updated as position changes to the memory model, and these updates are scheduled or triggered by the player or scene change. For decision-making, the LLM input combines memory model data with predefined prompts. LLM will output the content that needs to be executed in a specific format output that dictates Spirit interactions, animations, and vehicle generation based on participant engagement and appropriate difficulty levels. Finally, the output text is disassembled through the script and the corresponding animation or voice playback is executed.


\textbf{Synchronization of virtual and real space.} 
Within Unity, four cameras are set and orthographic mode is used to facilitate parallel projection. The views captured by these cameras are then mapped onto the projector's output using Spout, effectively translating the virtual scene into the physical space. Additionally, the positions of tracked participants are linked to a capsule object within Unity, allowing interaction between the players' locations and the colliders in the scene. To ensure accurate positional about the player's position within virtual and real space, the vectors tracked by the Tracker are multiplied by a coefficient \( n \) and adjusted by an offset \( b \). These two coefficient values will be manually calibrated every time the system starts.

\begin{figure*}
  \includegraphics[width=1\textwidth]{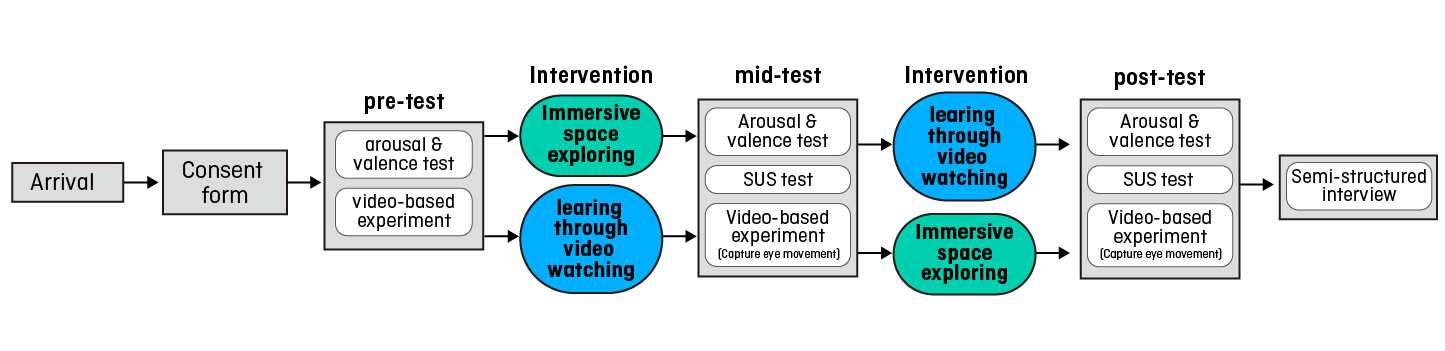}
  \caption{The experimental procedure of the study}
  \Description{This figure shows the experimental procedure of our study: Upon arrival, autistic children and their parents signed consent forms and were briefed on the experimental process. Children were randomly assigned to experimental groups and tested on social understanding and emotional responses. They then underwent either AIRoad or video-based learning interventions, followed by retesting with additional measures. Throughout, each child was accompanied by an experimenter.}
  \label{fig:procedure}
\end{figure*}


\begin{figure*}
 \includegraphics[width=0.8\textwidth]{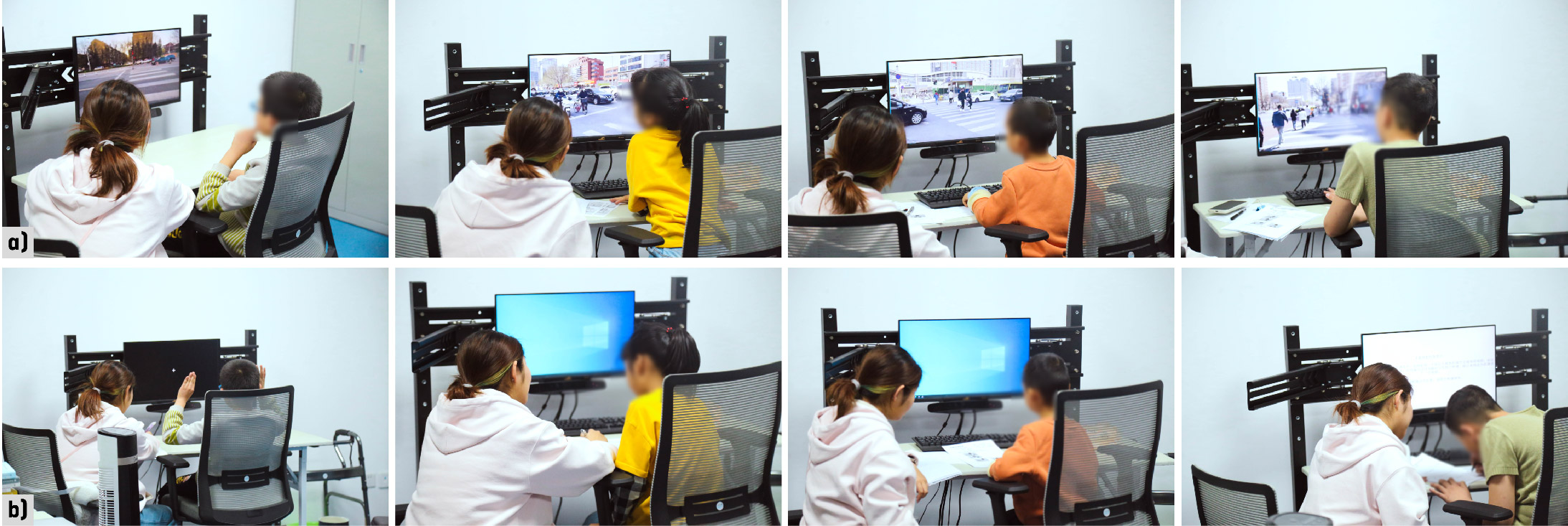}
 \caption{a) Children are participating in a video-based experiment; b) Completing a quiz on social affordance.}
 \Description{This series of images shows children and experimenters during the measurement phase of the experiment; a) Children are participating in a video-based experiment; b) Completing a quiz on social affordance.}
 \label{fig:experiment 1}
\end{figure*}

\begin{figure*}
 \includegraphics[width=0.8\textwidth]{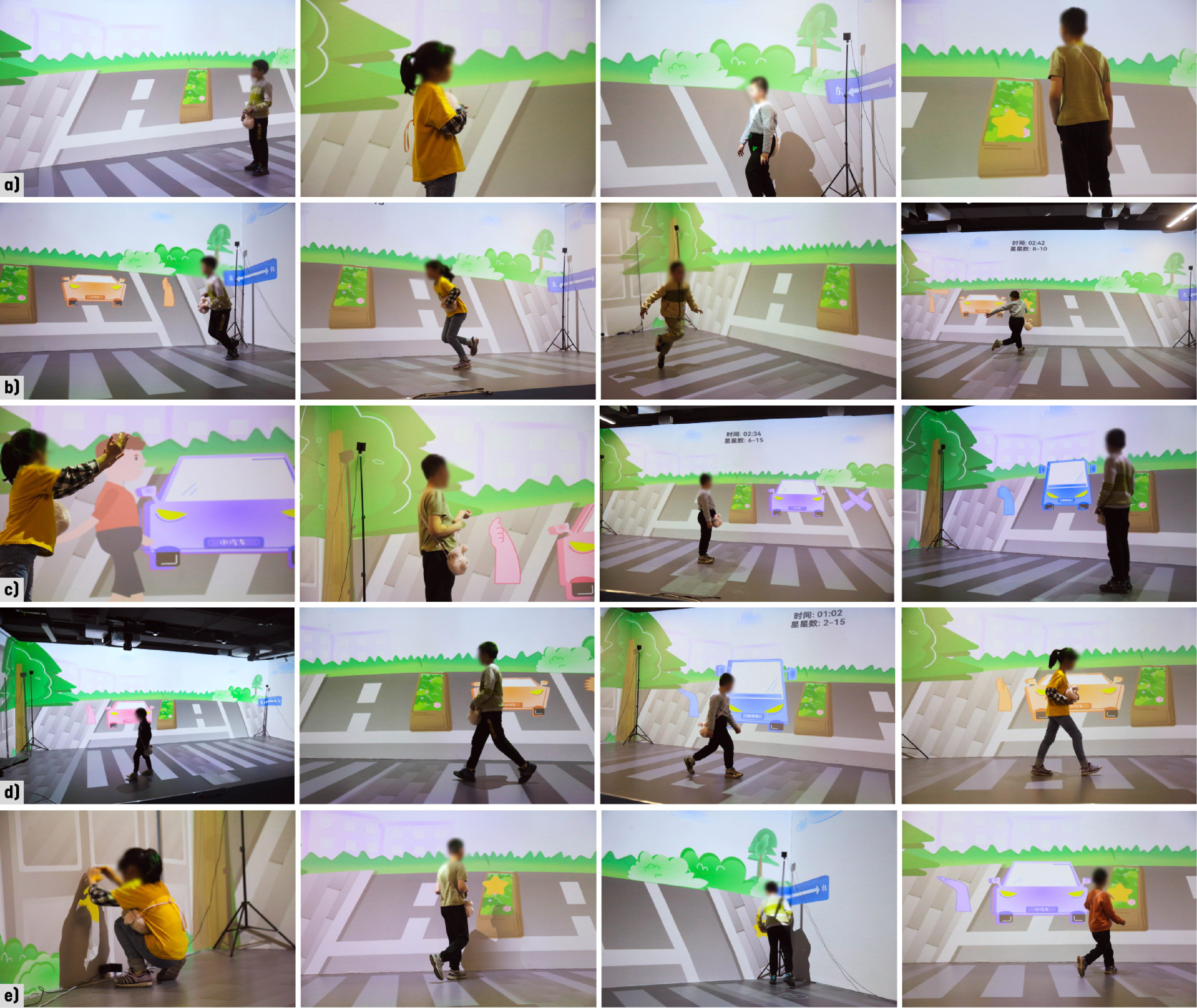}
 \caption{The procedure of immersive space exploring: a) Children observe their surroundings in AIRoad; b) Upon first seeing an approaching vehicle, they nervously run to the other side of the road; c) They observe and learn the different gestures and voice prompts of various vehicles; d) They gradually understand the intentions of different vehicles and begin to cross the road calmly; e) They successfully collect the stars and complete the mission.}
 \Description{This series of images illustrates the procedure of immersive space exploring: a) Children observe their surroundings in AIRoad; b) Upon first seeing an approaching vehicle, they nervously run to the other side of the road; c) They observe and learn the different gestures and voice prompts of various vehicles; d) They gradually understand the intentions of different vehicles and begin to cross the road calmly; e) They successfully collect the stars and complete the mission.}
 \label{fig:experiment 2}
\end{figure*}

\section{User Experiment}
After implementing the system, we conducted user experiments to evaluate its impact on real users. The primary research questions include the following three:

\begin{enumerate}
    \item[RQ1] To what extent does AIRoad demonstrate usability for autistic children? How do these children perform within the training?
    \item[RQ2] What are the engagement behaviors and emotional responses of autistic children when training with the AIRoad ?
    \item[RQ3] Can AIRoad effectively enhance the understanding of social affordances in traffic settings among autistic children?
\end{enumerate}



\subsection{Participants and Ethical Approval}
In our study, we recruited a total of 14 participants aged 6 to 12 years (M = 8.79, SD = 2.01), among whom 5 were recommended by a qualified special education institution, and the remaining 9 were recruited through a combination of online and offline methods. For the 9 participants, the recruitment process was structured and rigorous. Parents were required to provide self-reports on their children's conditions and complete the Autism Behavior Checklist (ABC). Additionally, parents voluntarily provided diagnostic reports from qualified institutions or hospitals, or professional opinions from certified specialists. This recruitment process enabled us to gather a suitable sample of participants while ensuring they met the necessary criteria for the study. This study was approved by the Ethics Committee of Tsinghua University.
\subsection{Experimental Design and Experimental Procedure}
We conducted a within-subject study comparing two experimental conditions: AIRoad training and video-based tutorials (Fig. \ref{fig:procedure}). To counterbalance potential order effects, participants were randomly divided into two groups, experiencing either AIRoad or the video-based tutorial first.
In the AIRoad condition, children first familiarized themselves with the virtual environment, then engaged in an immersive road scenario where they collected stars appearing in sequence (Fig. \ref{fig:experiment 2}). The task was complete upon reaching a predetermined number of stars. The interface displayed progress and remaining time to help children track their status. In the video condition, participants watched curated road safety videos~\footnote{https://www.youtube.com/watch?v=yEP3pws5lNQ\&t=1s}.
The experimental procedure consisted of three test sessions (pre-, mid-, and post-test), interspersed with two training processes, and concluded with a semi-structured interview. The procedure was as follows: (1) parent consent and briefing, (2) pre-test, (3) first training intervention (Fig. \ref{fig:experiment 1}), (4) mid-test using the same materials as pre-test, (5) parent completion of System Usability Scale (SUS) \cite{brooke1996sus}, given children's limited compliance in answering such questions, (6) second training condition, (7) post-test using the same materials as mid-test, and (8) concluding interview. Throughout the study, each child was assisted by two experimenters, with parents nearby to address any special circumstances. All sessions were video recorded.

\subsection{Measurement}
The study measurements consisted of three phases: test scores from pre-, mid-, and post-assessments, behavioral data logged by the AIRoad system, and post-hoc analysis of video recordings. This multi-faceted approach enabled thorough analysis of participants' learning progress and interactions.

\begin{enumerate}

\item \textbf{Video-Based experiments.} We recorded a set of videos depicting real street scenarios and asked autistic children to determine the appropriate moment to cross the street, aiming to evaluate participants' understanding of social affordance. Three authors collaboratively predetermined a safe period for each video, during which it was deemed safe to cross the street based on traffic light signals and passing vehicles. The accuracy and reaction time were computed according to these safe periods. A correct response was recorded if the child selected to cross the street within the predetermined safe period. The reaction time was measured as the duration between the start of the safe period and the participant's selection time.

\item \textbf{Level of Arousal and Valence.} To assess emotional changes in children before and after the training condition, we employed the Valence Arousal Scale, derived from Russell's circumplex model of affect. This model posits that emotion is composed of two bipolar and orthogonal dimensions~\cite{russell1980circumplex, russell1989affect}. Each scale ranged from -2 to 2, using integer values.

\item \textbf{SUS Scales.} To measure system usability, we employed an adapted version of the System Usability Scale (SUS) \cite{brooke1996sus} (see Appendix). Due to the limited compliance of autistic children in responding to such questionnaires, we asked parents to complete this scale on behalf of their children.

\item \textbf{In-game Log Data.} To evaluate autistic children's performance in the AIRoad training, we recorded road-crossing accuracy based on log data. Accuracy was computed through real-time position detection; a correct result was registered if the child was waiting or had arrived at a safe region when the car-leaving animation initiated. An incorrect result was recorded when the child collided with a vehicle.

\begin{table*}[htbp]
\centering
\caption{Manual coding scheme for Activity Participation and Emotional Activity}
\label{tab:my-table}
\begin{tabular}{ll}
\hline
Codes & Descriptions     \\ \hline
\multirow{2}{*}{Activity Participation} 
& \begin{tabular}[c]{@{}l@{}}Engrossed: Follow instructions, participate in activities according to rules or have a visual\\ gaze for major objects, 3 seconds is counted as 1 time \end{tabular} \\
& \begin{tabular}[c]{@{}l@{}}Distracted: Ignore instructions, avert or shift eyes to extraneous objects, and unable to \\ participate in activities according to rules\end{tabular}   \\ \hline
\multirow{2}{*}{Emotional Activity} 
& \begin{tabular}[c]{@{}l@{}}Emotional Expression: Exhibit associated emotions (pleasure, delight, agitation, excitement,\\ nervousness, etc.) \end{tabular} \\
& \begin{tabular}[c]{@{}l@{}}Problematic Behavior: Exhibit severe problem behavior (crying, emotional breakdowns, \\ aggressive behavior, etc.)\end{tabular}    \\  \hline
\end{tabular}
\end{table*}

\item \textbf{Video Data and Analysis.} The entire experiment was recorded using a GoPro and a 2K camera (as shown in Fig. \ref{fig:teaserCave}). These video recordings were primarily utilized for manual coding to analyze children's engagement and emotional expression \cite{1555162}. The coding scheme was developed based on and incorporated elements from Gong et al. \cite{gong2021holoboard} and Wu et al. \cite{wu2023mr}. All children's recordings were coded. Two trained coders independently coded these videos, achieving an inter-rater Pearson correlation greater than 0.99.

\end{enumerate}

\subsection{Results}
\subsubsection{\textbf{RQ1: System usability and participants' performance}}
The results of the \textbf{\textit{System Usability Scale (SUS)}} are summarized in Table \ref{tab:my-table}. The scores indicate that the usability of AIRoad was excellent (>85.0), acceptable (>70.0), and achieved an A+ level (>84.1) according to Bangor et al.'s empirical evaluation \cite{bangor2008empirical}. As evident from the data presented in Table \ref{tab:my-table}, AIRoad consistently scored higher than the video tutorial condition. Notably, AIRoad's score exceeded 85, placing it in the "Excellent" category.
These findings address RQ1, suggesting that parents of autistic children perceive AIRoad as an acceptable and highly usable tool for this population.

\begin{table}[ht]
\centering
\caption{System Usability Scale}
\label{tab:my-table}
\begin{tabular*}{\columnwidth}{@{\extracolsep{\fill}}lcc}
\hline
Experiments & AIRoad   & Video-watching \\ \hline
Mean        & 85.71    & 80.36          \\
SD          & 15.32    & 18.11          \\ \hline
\end{tabular*}
\end{table}

\begin{figure}[http]
  \includegraphics[width=0.4\textwidth]{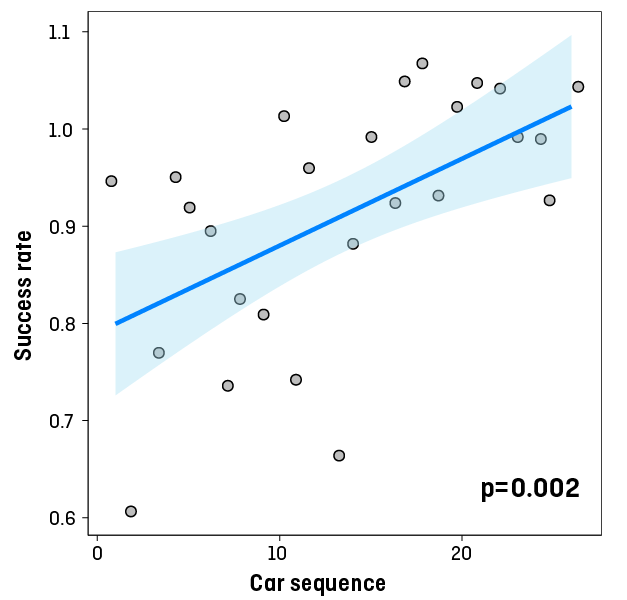}
  \caption{The relationship between the rate of pantcipants successfully crossing the road and the number of cars used in training.}
  \Description{This figure depicts the results of logging children's behavior in AIRoad, focusing on their rate of safely crossing the street. Logistic regression analysis revealed a significant gradual improvement in participants' ability to safely navigate the game environment (p=.002), demonstrating autistic children's capacity to learn safe strategies within the virtual space.}
  \label{fig:correct result}
\end{figure}

\textbf{\textit{Experimental training’s in-game performance}}
We conducted a logistic regression analysis to evaluate the behaviors of children in AIRoad, particularly their ability to safely cross the street. As shown in Fig.~\ref{fig:correct result}, the analysis revealed a significant correlation between the training process and the rate of safe crossings (p=0.002). This result addresses RQ1, indicating that through repeated experiences in the game training, autistic children were able to gradually reduce their error rates and demonstrate improved performance.
This finding may also suggest insights related to RQ3, indicating that autistic children may have gained a better understanding of social affordances. However, it is also possible that this improvement stems from their increasing familiarity with the game itself.

\begin{table*}[ht]
\caption{Children's t-test results of video coding}
\label{tab:CTVC}
\begin{tabular}{llllll}
\hline
Data & Variable & AIRoad & Video-watching & Value & P \\ \hline
\multirow{2}{*}{Activity Participation} & Engrossing & M=55.615(SD=2.980) & M=42.000(SD=18.478) & t=2.281 & 0.046* \\ 
& Distracting & M=4.385(SD=2.980) & M=18.000(SD=18.478) & t=-2.281 & 0.046* \\
\multirow{2}{*}{Emotional Activity} & Emotional Expression & M=1.539(SD=2.015) & M=0.227(SD=0.518) & t=4.719 & 0.001** \\ 
& Problematic Behaviour & M=5.500(SD=3.702) & M=2.546(SD=2.318) & t=-1.330 & 0.213 \\ \hline
\end{tabular}
\end{table*}


\subsubsection{\textbf{RQ2: Engagement and Emotional Responses in AIRoad Training}}
To examine the differences in children's engagement and emotional experiences between AIRoad and Video-watching, we employed a paired-sample T-test for video coding analysis. Analysis of the\textbf{\textit{Video coding}} from the children's tests, as presented in Table \ref{tab:CTVC}, indicates that participants in the AIRoad condition demonstrated higher levels of activity participation and emotional engagement. 

Regarding the Activity Participation Results, the mean of engrossing behavior in AIRoad condition was 55.615 (SD = 2.980) and of engrossing in Video-tutorial condition was 42.000 (SD = 18.478). This difference was statistically significant according to a paired-samples t-test (t(13) = 2.281, p < .05). 
And the mean of distracting behavior in AIRoad condition was 4.385 (SD = 2.980) and of distracting behavior in Video-tutorial condition was 18.000 (SD = 18.478). This difference was statistically significant according to a paired-samples t-test (t(13) = -2.281, p < .05).

In terms of Emotional Activity Results, the mean of emotional expressions in AIRoad condition was 1.539 (SD = 2.015) and of emotional expressions in Video-tutorial condition was 0.227 (SD = 0.518). This difference was statistically significant according to a paired-samples t-test (t(13) = 4.719, p < .001). 
And the mean of problematic behavior in AIRoad condition was 5.500 (SD = 3.702) and of problematic behavior in Video-tutorial condition was 2.546 (SD = 2.318). This difference did not show statistically significant according to a paired-samples t-test (t(13) = -1.330, p > .05).

These findings suggest that children exhibit more positive affective states and increased attentional focus when engaged in AIRoad training than video tutorial.
These results directly address RQ2, providing evidence that the AIRoad system we developed can offer autistic children a more engaged experience and promote positive emotional states.

\begin{figure}
  \includegraphics[width=0.5\textwidth]{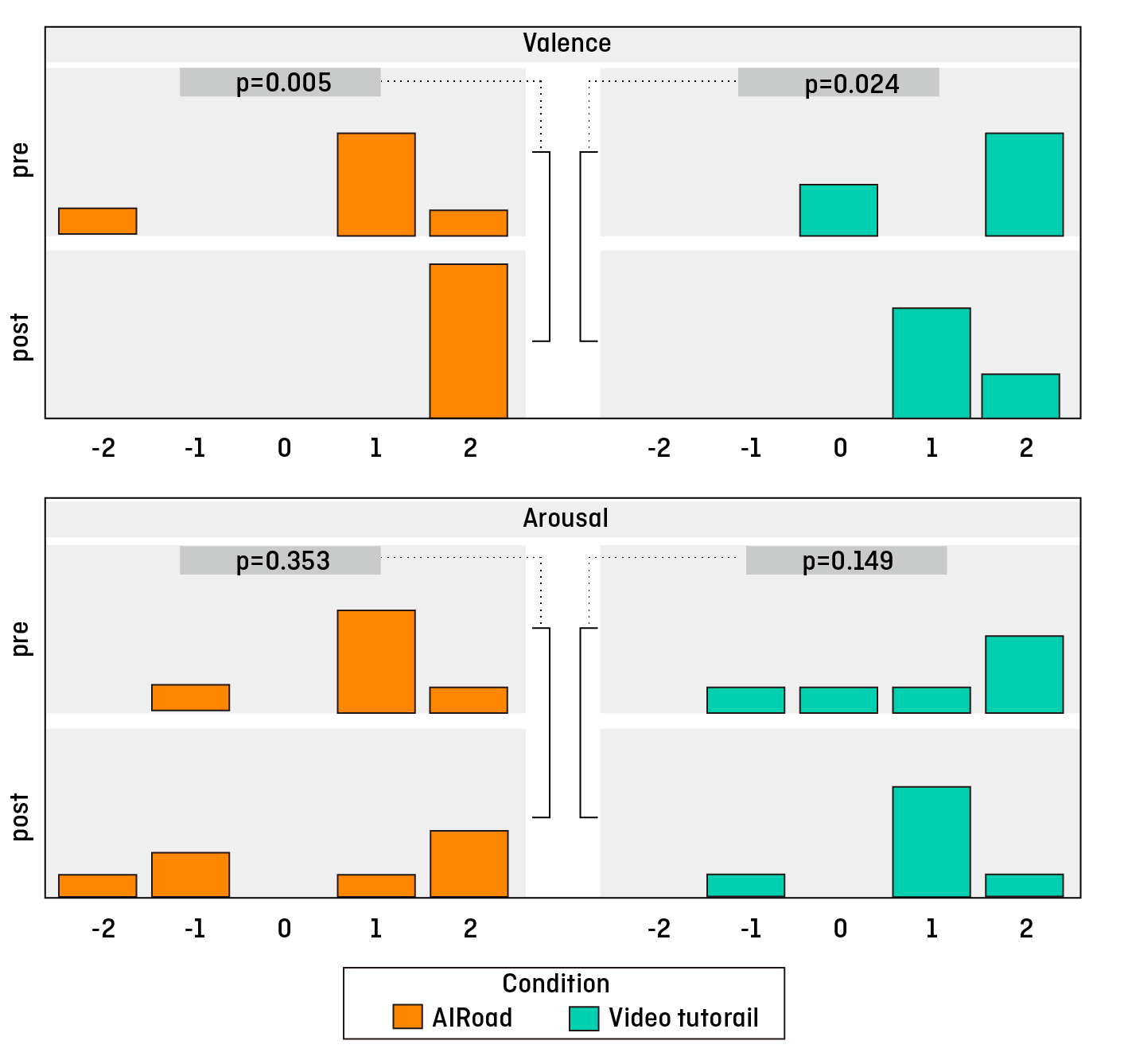}
  \caption{The result of Valence and Arousal Scale.}
  \Description{This figure shows the results of the Valence and Arousal Scale, indicating a significant improvement in children's valence following AIRoad training (p=.005), while a notable decrease was observed after video-based learning (p=.024). The arousal levels did not show any significant changes.}
  \label{valence and arousal}
\end{figure}

We conducted a statistical analysis of the results from the\textbf{ Valence and Arousal Scale}.
As shown in Fig. \ref{valence and arousal}, we conducted the Fisher test on the Valence and Arousal Scale. The results reveal a significant improvement in children's valence after undergoing AIRoad training (p<.01). Conversely, following video-watching educational learning, a significant decrease in children's valence was observed (p<.05). In contrast, no significant changes in arousal levels were detected in autistic children before and after the training with either educational technology (p = .353 / .149).

This result addresses RQ2, indicating that AIRoad can bring positive emotions to autistic children, although there are no significant differences in the intensity of these emotions. Meanwhile, traditional video tutorials may lead to negative emotions in autistic children.

\subsubsection{\textbf{RQ3: Effects on Enhancing the Understanding of Social Affordances}}
\begin{figure}[htbp]
  \includegraphics[width=0.45\textwidth]{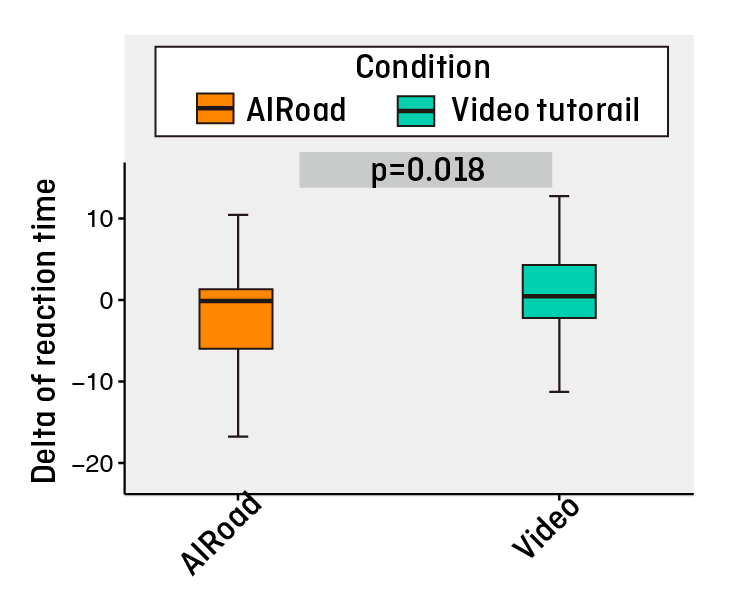}
  \caption{Reaction Time}
  \Description{This figure shows the changes in response time of autistic children to road videos between AIroad training and video tutorial groups. The AIRoad group demonstrated significantly lower response times, p=0.018.}
  \label{fig:Reaction}
\end{figure}

\textbf{\textit{Video-based test}}
To assess the impact of training on autistic children's response times to video stimuli, we calculated the difference in reaction times before and after the training intervention. This measure is referred to as the \textit{Delta of Reaction Time}. 
The result of \textit{Delta of Reaction Time} can be seen in Fig. \ref{fig:Reaction}. 
The mean \textit{Delta of Reaction Time} for the AIRoad condition was -2.20 (SD = 6.67), while for the video tutorial condition it was 0.18 (SD = 6.00). The Wilcoxon rank-sum test indicated a statistically significant difference between the two groups (W = 2343, \textit{p} < .05), suggesting that the \textit{Delta of Reaction Time} in the video tutorial condition was significantly higher than in the AIRoad condition.
These findings indicate that after AIRoad training, participants were able to respond more quickly to the appropriate time to cross the street. This also partially addresses RQ3, suggesting that autistic children exhibited a better understanding of social affordance in the context of the street-crossing scenario following their experience with AIRoad. Participant 3 noted, \textit{"But some cars, even though they're far away, they drive so fast and zoom; they're here in no time. So, just because they're far doesn't mean it's safe to go... And if they're close, it's not always okay; I gotta see if they're speeding fast or not."} Participant 7 also mentioned, \textit{"If a car's going fast, it means they're in a hurry for something important, like going to work or to the hospital... And if a car's going slow, it means they don't have anything important going on."} These qualitative results further demonstrate that autistic children gradually connected vehicle behavior with its underlying intentions through repeated practice in AIRoad training, enhancing their understanding of social affordance within traffic settings.

\section{Discussion}

\subsection{Large Language Models as Special Educational Support Tools for Autistic Children}

LLMs have shown utility beyond common applications, as this study explores their potential in special education for autistic children.
LLMs have shown utility beyond common applications, as this study explores their potential in special education for autistic children. Leveraging autonomous decision-making and social simulation capabilities, LLMs offer promising solutions through consistent and adaptable learning tools, reducing burdens on educators.
Their adaptability enables customized feedback and guidance based on real-time interactions~\cite{lyu2024designing}. Moreover, their social simulation capabilities create safe, controlled environments for autistic children to practice social skills, generating unlimited possible social scenarios through their generative abilities.
This aligns with previous research demonstrating that virtual environments provide secure spaces for children to rehearse social behaviors without real-world unpredictability~\cite{matsentidou2014immersive}. By generating dynamic, contextually relevant dialogues, LLMs enhance these virtual environments, enabling practice across diverse social scenarios.
While direct real-world experience remains essential, LLMs serve as valuable preparatory tools. The ability to simulate social interactions in virtual settings facilitates repeated practice—crucial for autistic children who often struggle with skill generalization. Research has shown that children who engage with virtual environments for social skills training successfully transfer these skills to real-world contexts~\cite{matsentidou2014immersive, paneru2024nexus}. 
For LLMs, hallucination remains one of their most significant risks.
In our evaluation of 100 simulated LLM responses, we identified 6 hallucination cases (6\%): 2 instances of extraneous text generation and 4 cases of intention-behavior inconsistencies. While Unity's processing system filters extraneous text, intention-behavior contradictions are addressed by introducing the concept of "lying cars" to autistic children during experiments. 
Although the 6 \% hallucination rate is relatively low, future research could employ self-reflection techniques to further reduce these occurrences~\cite{ji2023towards}. Given the rapid advancement of large language models, subsequent work could explore more advanced models to enable more immediate and diverse interactions.

\begin{figure*}[htbp]
  \includegraphics[width=1.0\textwidth]{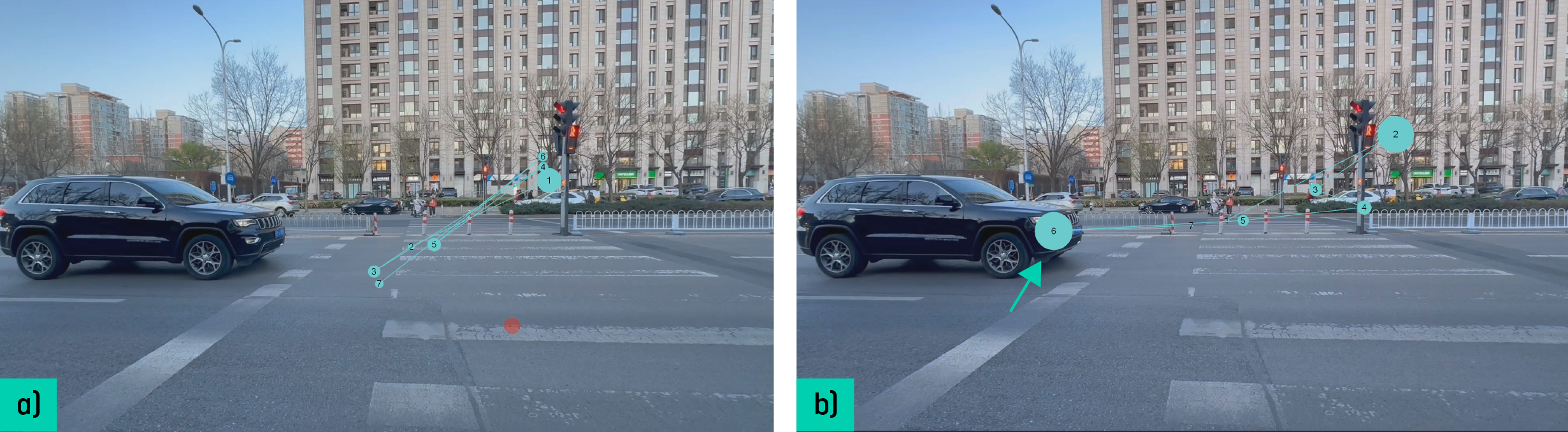}
  \caption{Case on the Changes in Eye Movement Before and After Training with AIRoad. a) Eye movement case before AIRoad training
b) Eye movement case after AIRoad training.}
  \Description{This figure shows a case study on the changes in eye movement before and after training with AIRoad. After AIroad training, autistic children focused their visual attention more on vehicles.}
  \label{eyetrack}
\end{figure*}

\subsection{Benefits and Experiences of Immersive Interfaces for Autistic Children}

Projection-based immersive spaces demonstrate significant advantages in social simulation for autistic children. The medium system proves accessible and acceptable to autistic children. Embodied learning in spatial environments provides children with enhanced immersion and improved emotional engagement, addressing attention deficits and learning challenges common among autistic children~\cite{mcgee2001educating}. Their progressively improving performance demonstrates the effectiveness of learning social affordance in virtual scenarios. Simulation serves as a vital bridge before real traffic exposure for autistic children who typically find direct real-world learning challenging - a proven method in special education~\cite{rao2008social}. Although in virtual spaces, our ultimate goal is to enable autistic children to perform better in the real world through virtual interactions. AIRoad shows promising real-world skill transfer potential, supported by improved response times to real road footage.
Moreover, the decision-making processes of autistic children in the immersive AIRoad environment closely mirrored those observed in video-based tests. For instance, Participant 3 displayed risky behavior, attempting to run across the road during AIRoad training, a behavior echoed in the video-based test. In contrast, Participant 1 consistently waited for vehicles to pass completely before crossing. This suggests that immersive environments can effectively expose the real challenges faced by autistic children in a safe context. Further exploration of rehabilitation effects under simpler equipment conditions is possible~\cite{cosentino2023moves}, as more accessible interactive projection systems have been developed.

\subsection{Autistic children became more attentive to the characteristics of cars after training in AIRoad}

One possible explanation for the changes in response times lies in the shifts of attention patterns, as revealed by our eye-tracking data during video response tests. 
After the training, autistic children not only pay attention to traffic lights but also begin to observe vehicles and their adherence to traffic rules. Prior to AIRoad training, most children understood only the explicit rules for crossing the street—such as stopping at a red light, walking on a green light, and waiting for cars to pass before proceeding. However, these rules often fall short in traffic situations that are rich with complex social signals.
The training resulted in a noticeable shift in children's attention patterns towards vehicle signals. This shift is evidenced in eye movement patterns, as shown in Fig.\ref{eyetrack}. The data reveals that post-training, children allocated more time observing vehicles. While participants initially focused primarily on traffic lights and road conditions, after exploring AIRoad, they redistributed their attention to include parked vehicles.
However, this modified attention pattern occasionally led to misjudgments. For example, P11, who previously relied on traffic lights for crossing decisions, began focusing more on vehicle movements post-training, assuming it was safe to cross whenever cars stopped. As P11 noted, \textit{"In the future, I will pay more attention when crossing the road, and I will look out for the driver's hand gestures in the cars."} It is necessary to generate more diverse scenarios to help autistic children integrate multiple cues in their decision-making process.

\section{Limitation and future work}
Currently, AIRoad still faces numerous limitations. For example, the hallucinations of LLMs mentioned above are difficult to control \cite{xu2024hallucination}, leading to the generation of unexpected content in speech and the simulation of affordances, thereby increasing the system's unpredictability. Additionally, the latency of LLMs to some extent affects the system's response time. Our system currently mitigates the impact of delays by generating multiple vehicles simultaneously, but future upgrades and miniaturization of LLMs may potentially improve this issue \cite{shi2023research}.

Secondly, in the future, it is possible to construct more social scenarios filled with complex signals to simulate the social affordances they comprise, such as richer traffic scenarios including public transportation, among others. Moreover, there could be attempts to simulate a wider range of daily life scenarios. Wherever there is a human environment, the presence of social affordances is inevitable \cite{loveland1991social}.
Currently, the primary input modality for children is positional information. In the future, it's worth considering interactions through additional modalities, such as vocal input or even tactile interaction. With a richer array of interaction modalities in the space, children might experience a stronger sense of immersion and more effective learning outcomes \cite{anastopoulou2004investigating}.

\section{Conclusion}
This study first conducted a comprehensive review of educational technologies for autistic children, resulting in the identification of 17 design guidelines across four major categories, which laid the foundation for our LLM-enabled immersive system. Using a street-crossing scenario, the system simulated diverse driving behaviors through multimodal signals, including visual cues and auditory prompts. This provided autistic children with a safe environment to practice interpreting and responding to social signals. A user study involving 14 autistic children demonstrated significant improvements in their decision-making processes, suggesting an enhanced understanding of social affordances in traffic settings. Our system effectively enhances autistic children's ability to recognize and interpret social affordances by providing a controlled yet dynamic environment where they can safely explore and respond to complex social cues. Promising future research could investigate expanding this system across a broader range of social contexts, such as classroom or public space interactions, and explore the long-term transfer of learned skills to real-world settings through longitudinal studies.

 \section{Acknowledgments}
We would like to thank Shanghai Children's Welfare Institute for their inspiration and assistance in this research. This work was supported by the National Natural Science Foundation of China Youth Fund (No. 62202267).

 \bibliographystyle{ACM-Reference-Format}
 \bibliography{sample-base}

\appendix

\section{Experimental Details}

\subsection{Online questionaire}
\textbf{Investigation on ASD children's language and motor skills.}\\
Hello,Parents!Thank you very much for participating in this questionnaire. We have designed a virtual space environment for autistic children to explore, where they can learn and interact.\\
All your answers are for statistical analysis and academic research only. Please fill in the questionnaire carefully according to the actual situation of your child and feel free to answer! If you and your child are interested in participating in our "Little Explorer" virtual space experience and becoming our little explorers, please leave your contact information at the end of the questionnaire.

Your child's name:
What is your child's gender?
\begin{itemize}
    \item male
    \item female
\end{itemize}

Please enter your child's date of birth:

How is your child's motor skills?
\begin{itemize}
\item Very good
\item Good
\item Average
\item Poor
\item Very poor
\end{itemize}

How is your child's language skills?
\begin{itemize}
\item Very good
\item Good
\item Average
\item Poor
\item Very poor
\end{itemize}

Your child's personality:
\begin{itemize}
\item Outgoing and lively
\item Introverted and quiet
\item Curious and exploratory
\item Sensitive and sentimental
\end{itemize}

Does your child enjoy going on trips?
\begin{itemize}
\item Like very much
\item Not so much
\item Neutral
\item Dislike
\item Very Dislike
\end{itemize}

What is your usual mode of trips?
\begin{itemize}
\item Self-drive
\item Public Transport
\item Walking
\item Cycling
\item Other
\end{itemize}

Has your child travelled independently?
\begin{itemize}
\item Yes, he/she can travel fully independently
\item limited independent trips, need some guidance
\item cannot travel independently, always needs to be accompanied
\item never tried to travel independently
\end{itemize}

\section{Adapted SUS scale}
(five-Likert scale,1:It fits perfectly;5:Very inconsistent)
\subsection{AIRoad}
\begin{itemize}
\item I think I would let my child play this game often
\item I think this game is too difficult for children
\item I think this game is too easy for the child to play
\item The child needs help to complete the game
\item I think the system is well integrated with various functions
\item I think other children would learn the game very quickly as well
\item I think there are too many inconsistencies in the system
\item I found the system very cumbersome to use
\item I think the system is well designed and fun for children to play 
\item The child would need to know a lot of other knowledge and information beforehand to play the game
\end{itemize}

\subsection{Video-watching}
\begin{itemize}
\item I think I would let my child watch these videos often
\item I think this video is too difficult for children
\item I think this video is too easy for my child to watch
\item My child needs help to watch this video independently
\item I think this video has a good integration of various elements
\item I think other children would learn the content of this video very quickly as well
\item I think there are too many inconsistencies in the video
\item I found it very cumbersome to watch and use the video
\item I think the video is well designed and interesting for children to watch
\item Children need to know a lot of other knowledge and information before they can watch this video
\end{itemize}










\end{document}